\documentclass{pic2012}
\def\runauthor{Roberto~Salerno on behalf of the CMS Collaboration}
\def\shorttitle{Higgs searches at CMS}
\usepackage{subfig}
\begin{document}

\title{Higgs searches at CMS}

\author{Roberto~Salerno}

\address{
{\rm On behalf of the CMS Collaboration}\\
CNRS-IN2P3 Laboratoire Leprince-Ringuet\\
Ecole polytechnique\\
91128 PALAISEAU Cedex, France\\
E-mail: roberto.salerno@cern.ch 
}

\maketitle

\abstracts{
Results are summarized from searches for the standard model Higgs boson in $pp$ collisions
at $\sqrt{s} = 7$ and 8 TeV in the CMS experiment at the LHC.
The measurements of mass, cross section, and properties of the narrow 
resonance recently observed are presented.
The searches beyond standard model Higgs boson, in the CMS experiment at the LHC,
are highlighted.
}

\section{Introduction}

Earlier this year the ATLAS and CMS collaborations announced the 
observation of a narrow resonance with mass near 125 GeV~\cite{Aad:2012zz,Chatrchyan:2012zz} with properties consistent 
with that of the Higgs boson predicted in the standard model (SM) of particle physics~\cite{StandardModel67_1,StandardModel67_2,StandardModel67_3}.
After a short review of the CMS detector,
we summarize 
the results from searches for the SM Higgs boson in $pp$ collisions
at $\sqrt{s} = 7$ and 8 TeV in the CMS experiment at the LHC and 
the measurements of mass, cross section, and properties of the narrow 
resonance recently observed. 
The searches beyond SM Higgs boson are highlighted.
%%
%%
%%
%%the status of the CMS 
%%searches for the standard model Higgs boson and the 
%%measurements of mass, cross section, and properties of the narrow 
%%resonance recently observed.
%%The CMS searches beyond standard model Higgs boson
%%are highlighted.
%%

\section{CMS detector and event reconstruction}

%%Particles produced in the $pp$ collisions are detected in the pseudorapidity range $ | \eta | < 5$,
%%where .
The CMS detector comprises a superconducting solenoid, providing a uniform magnetic
field of 3.8 T in the bore, equipped with silicon pixel and strip tracking systems  ($ | \eta | < 2.5$
\footnote{where $\eta = - \ln[\tan (\theta / 2)]$ and $\theta$ is the polar angle with respect to the direction
of the proton beam}
)
surrounded by a lead tungstate crystal electromagnetic  calorimeter (ECAL) and a brass-scintillator
hadronic calorimeter (HCAL) ($ | \eta | < 3.0$).
A steel/quartz-fiber Cherenkov calorimeter extends the coverage ($ | \eta | < 5$).
The steel return yoke outside the solenoid is instrumented with gas ionization detectors used to
identify muons ($ | \eta | < 2.4$).
A detailed description of the detector is given in Ref.~\cite{Chatrchyan:2008zzk}.

The CMS ``particle-flow'' event description algorithm~\cite{CMS-PAS-PFT-10-001,CMS-PAS-PFT-10-003}
is used to reconstruct and identify each single particle with an
optimized combination of all subdetector information.
In this process, the identification of the particle (photon,
electron, muon, charged hadron, neutral hadron) plays an important
role in the determination of the particle momentum. 
%The reconstructed particles are henceforth referred to as objects.

%\section{Event reconstruction}

\section{Searches for the standard model Higgs boson}

Four main mechanisms are predicted for Higgs boson production in $pp$ 
collisions:
the gluon-gluon fusion mechanism, which
has the largest cross section,
followed in turn by vector-boson fusion (VBF),
associated $WH$ and $ZH$ production ($VH$),
and production in association with top quarks ($t\bar{t} H$).

The particular set of sensitive
decay modes of the SM Higgs boson depends strongly on $m_H$.
The results presented here are based on the five most sensitive
decay modes:
$H \to \gamma\gamma$;
$H \to ZZ$ followed by $ZZ$ decays to $4\ell$;
$H \to WW$ followed by decays to $2\ell 2\nu$;
$H \to bb$ followed by $b$-quark fragmentation into jets;
and $H \to \tau\tau$ followed by at least one leptonic $\tau$ decay.
This list is presented in Table~\ref{tab:chans} and comprises the full
set of decay modes and subchannels, or categories, for which both the 7~and
8 TeV data sets have been analysed at the time of PIC 2012 conference.
New preliminary results, analyzing more integrated luminosity, have been recently produced by 
the CMS collaboration~\cite{CMS-PAS-HIG-12-041,CMS-PAS-HIG-12-042,CMS-PAS-HIG-12-043,CMS-PAS-HIG-12-044,CMS-PAS-HIG-12-045}, 
they are not included in this proceeding.

\begin{table*}[htbp]
  \begin{center}
    \caption{Summary of the subchannels, or categories, used in the
      analysis of each decay mode.}
    \label{tab:chans}
    \begin{tabular}{l c c c c} \hline
Decay & Production & No. of  & \multicolumn{2}{c}{Int.\ Lum.\ ($fb^{-1}$)} \\
mode  & tagging & subchannels  & 7 TeV & 8 TeV \\ 
\hline\hline
{$\gamma\gamma$} & untagged & 4 & {5.1} & {5.3} \\
                                         & dijet (VBF) & 1 or 2 & & \\ \hline
$ZZ$ & untagged & 3  & 5.1 & 5.3 \\ \hline
$WW$ & untagged & 4  & {4.9} & {5.1} \\
                                           & dijet (VBF) & 1 or 2 & & \\ \hline
$\tau\tau$ & untagged & 16  & {4.9} & {5.1} \\
                                          & dijet (VBF) & 4 & & \\ \hline
$bb$ & lepton, $E_T^{miss}$ (VH) & 10  & 5.0 & 5.1 \\ \hline
    \end{tabular}
  \end{center}
\end{table*}

For a given value of $m_H$, the search sensitivity depends on
the production cross section, the decay branching fraction into the chosen final state,
the signal selection efficiency, the mass resolution,
and the level of background from identical or similar final-state topologies.
For low values of the Higgs boson mass, the $H \to \gamma\gamma$ and $H \to ZZ \to 4\ell$ 
channels play a special role due to the excellent mass resolution for the reconstructed diphoton and four-lepton final states, 
respectively. 
The $H \to WW \to \ell\nu\ell\nu$ channel provides high sensitivity but has relatively poor mass resolution due to the presence of neutrinos in the final state. 
The sensitivity in the $bb$ and $\tau\tau$ decay modes is reduced due to the large backgrounds and poor mass resolutions. 
In the high mass range, the sensitivity is driven by the $WW$ and $ZZ$ modes.
In the following subsections a brief description of 
each decay modes is presented.

\subsection{$H \to \gamma\gamma$ decay mode} 

In the $H \to \gamma\gamma$ analysis~\cite{Chatrchyan:2012zz,Chatrchyan:2012tw} a search is made for a narrow peak in the diphoton
invariant mass distribution in the range 110--150 GeV,
on a large irreducible background from QCD production of two photons.
There is also a reducible background where one or more
of the reconstructed photon candidates originate from
misidentification of jet fragments.

To enhance the sensitivity of the analysis, candidate diphoton events are separated into mutually
exclusive categories of different expected signal-to-background ratios,
based on the properties of the reconstructed photons and on
the presence of two jets satisfying criteria aimed at selecting events
in which a Higgs boson is produced through the VBF process.
The analysis uses multivariate techniques
for the selection and classification of the events.

The background is estimated from data, without the use of MC
simulation, by fitting the diphoton invariant mass distribution in each of the categories
in a range (100 $ < m_{\gamma\gamma} < $ 180 GeV) extending slightly above and
below that in which the search is performed.
The choices of the function
used to model the background and of the fit range are made based on a study of the possible
bias in the measured signal strength.
Polynomial functions are used.

\begin{figure}[htbp]
 \subfloat[]{%
	\begin{minipage}[c][1\width]{%
	   0.5\textwidth}
	   \centering%	
	   \includegraphics[width=1\linewidth]{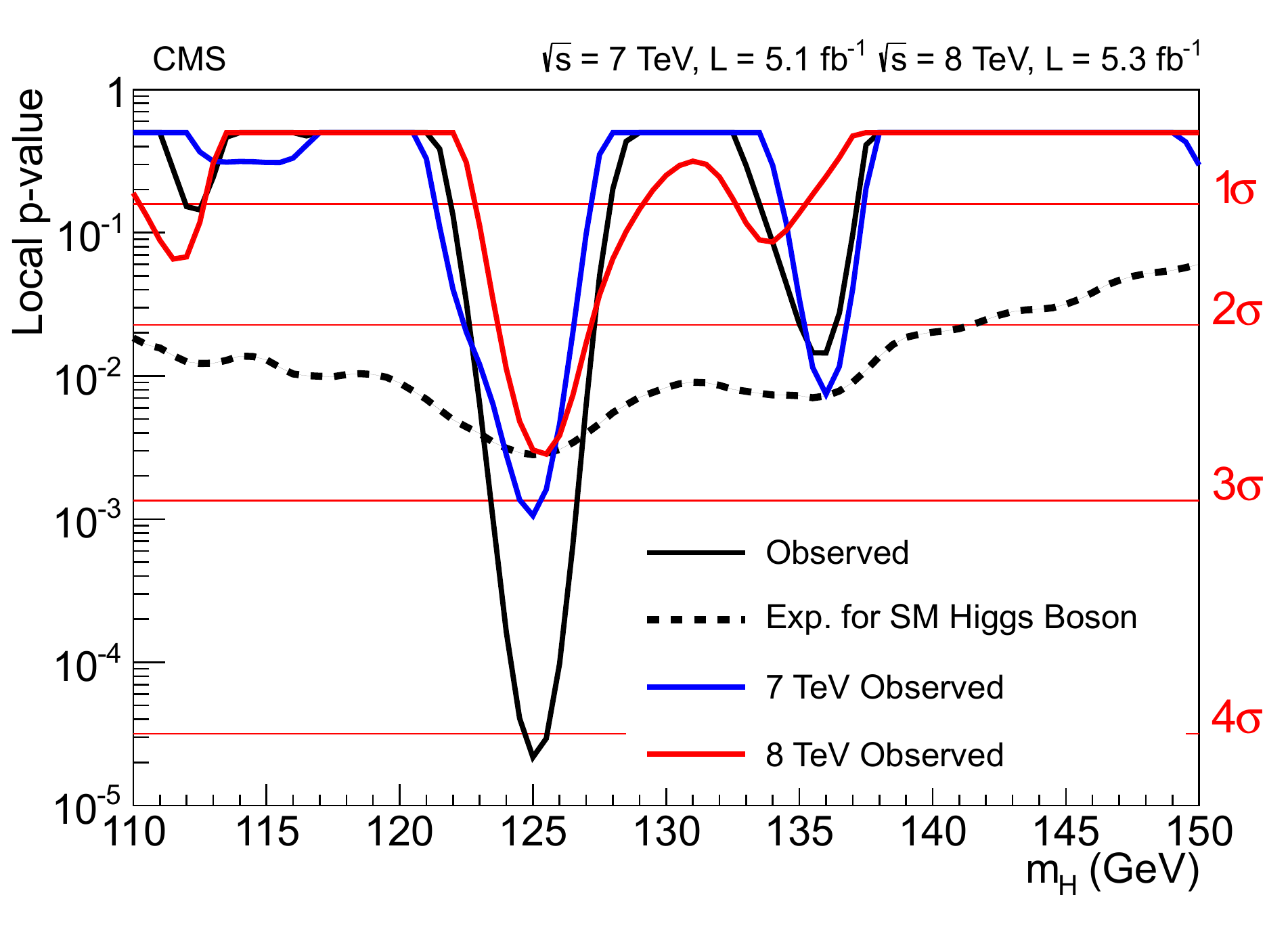}
	\end{minipage}}
	 \subfloat[]{%
	\begin{minipage}[c][1\width]{%
	   0.5\textwidth}
	   \centering%	
	   \includegraphics[width=1\textwidth]{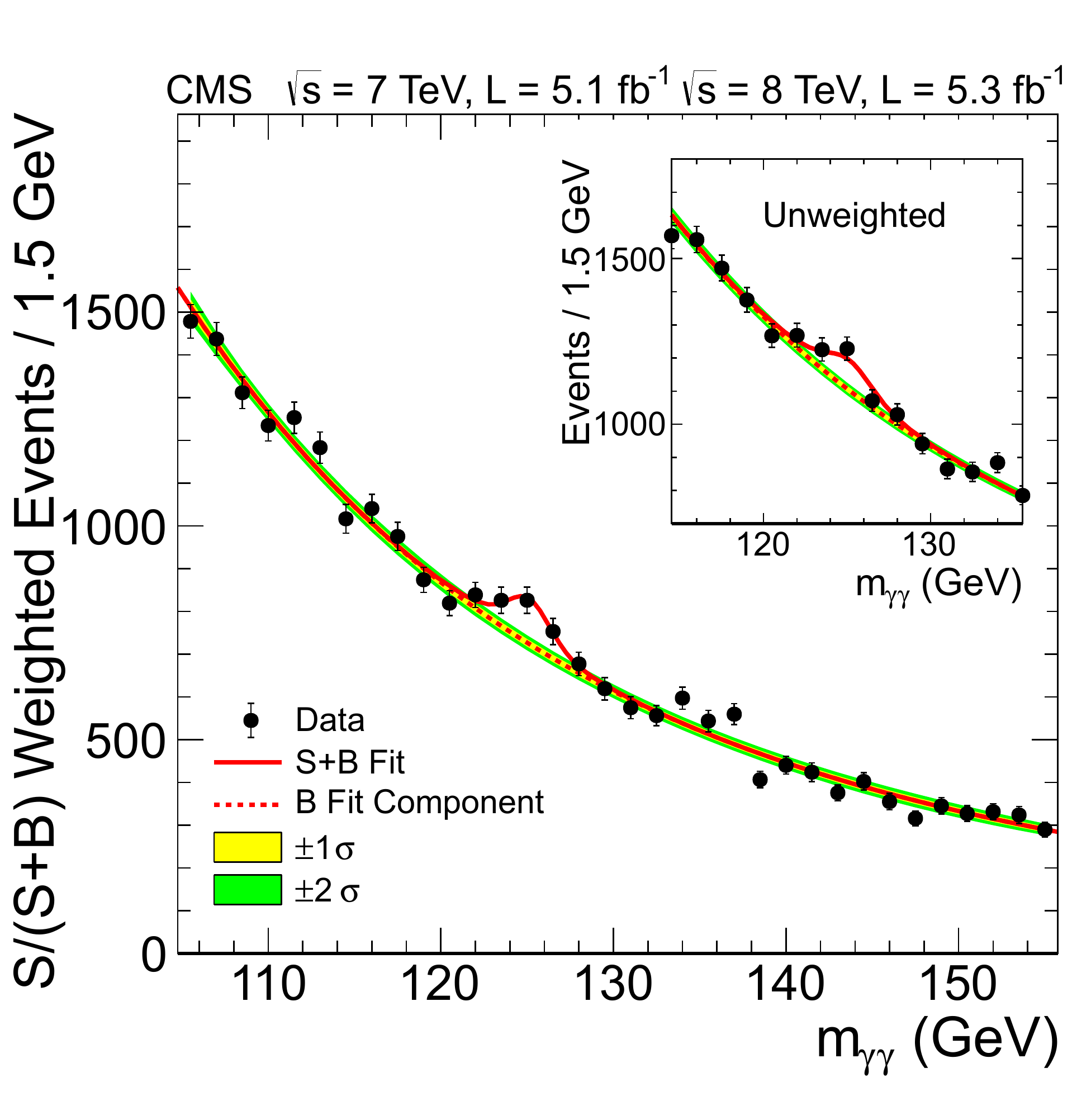}
	\end{minipage}}
     \caption{\small{
       (a) The local $p$-value as a function
       of $m_H$ in the $\gamma\gamma$ decay mode for the combined 7~and 8 TeV data sets. 
       The additional lines show the values for the two data sets taken
       individually. The dashed line shows the expected local $p$-value for the combined
       data sets, should a SM Higgs boson exist with mass $m_H$.
       (b) The diphoton invariant mass
       distribution with each event weighted
       by the $S/(S+B)$ value of its category. The lines
       represent the fitted background and signal, and the coloured
       bands represent the $\pm$1 and $\pm$2 standard deviation
       uncertainties in the background estimate.
       The inset shows the central part of the unweighted invariant
       mass distribution.
}}
\label{fig:gg}
\end{figure}

The expected 95\% CL upper limit on the signal strength
$\sigma/\sigma_\mathrm{SM}$, in the background-only hypothesis,
for the combined 7~and 8 TeV
data, is less than 1.0 in the range
110 $< m_H <$ 140 GeV, with a value of 0.76 at
$m_H$ = 125 GeV.
The observed limit indicates the presence of a significant excess at $m_H$ =
125 GeV in both the 7~and 8 TeV data.
%The features of the observed limit are confirmed
%by the independent sideband-background-model and cross-check analyses.
The local $p$-value is shown as a function of $m_H$ in
Figure~\ref{fig:gg}(a) for the 7~and 8 TeV data,
and for their combination.
The expected (observed) local $p$-value for a SM Higgs boson of mass 125 GeV corresponds 
to $2.8\,(4.1)\,\sigma.$ 
%In the sideband-background-model and cross-check analyses, the observed local
%$p$-values for $m_H=125$ GeV correspond to 4.6 and $3.7\,\sigma$, respectively.
The best-fit signal strength for a SM Higgs boson mass hypothesis of
125 GeV is $\sigma/\sigma_\mathrm{SM} = 1.6\pm0.4$.

In order to illustrate, in the $m_{\gamma\gamma}$ distribution, 
the significance given by the statistical methods, it is necessary to take into account the large
differences in the expected signal-to-background ratios of the event
categories.
The events are weighted according to the category in which they fall.
A weight proportional to $S/(S+B)$ is used
where $S$ and $B$ are the number of signal and background events.
Figure~\ref{fig:gg}(b) shows the data, the signal
model, and the background model, all weighted.
The unweighted distribution, using the same binning but in a more
restricted mass range, is shown as an inset.
The excess at 125 GeV is evident in both the weighted and unweighted distributions.

\subsection{$H \to ZZ\to 4\ell$ decay mode} 

In the $H \to ZZ\to 4\ell$ decay mode~\cite{Chatrchyan:2012zz,Chatrchyan:2012dg} a search is made for a narrow four-lepton mass peak
in the presence of a small continuum background.
Since there are differences in the reducible background rates 
and mass resolutions between the subchannels
4e, 4$\mu$, and 2e2$\mu$, they are analysed separately.
The background sources include
an irreducible four-lepton contribution from direct $ZZ$
production via $qq$ and gluon-gluon processes.
Reducible contributions arise from
$Z+b\bar{b}$ and $t\bar{t}$ production where the final states contain
two isolated leptons and two $b$-quark jets producing secondary
leptons.
Additional background arises from $Z+$jets and $WZ+$jets events 
where jets are misidentified as leptons.

The event selection requires two pairs of same-flavour, oppositely charged
leptons.
The pair with invariant mass closest to the Z boson mass is required to have a mass in the
range 40--120 GeV and the other pair is required to have
a mass in the range 12--120 GeV.
Electrons are required to have $p_T > 7$ GeV and $|\eta| < 2.5$.
The corresponding requirements for muons are $p_T> 5$ GeV
and $|\eta| < 2.4$.
Both muons and electrons are required to be pass identification requirements, 
based on a multivariate techniques, and to be isolated.
The electron or muon pairs from $Z$ boson decays are required to originate
from the same primary vertex.
Final-state radiation from the leptons is recovered and included
in the computation of the lepton-pair invariant mass.
The $ZZ$ background is evaluated from MC simulation studies.
The reducible and instrumental backgrounds are estimated from data 
selecting events in a background control region, well
separated from the signal region,
by relaxing the isolation and identification criteria for two
same-flavour reconstructed leptons.
The combined signal reconstruction and selection efficiency,
with respect to the $m_H = 125$ GeV generated signal with  $m_{\ell\ell} > 1$ GeV as the only cut, 
is 18\% for the 4e channel, 40\% for the 4$\mu$ channel, and 27\% for the 2e2$\mu$ channel.

\begin{figure}[htbp]
 \subfloat[]{%
	\begin{minipage}[c][1\width]{%
	   0.5\textwidth}
	   \centering%	
	   \includegraphics[width=1\textwidth]{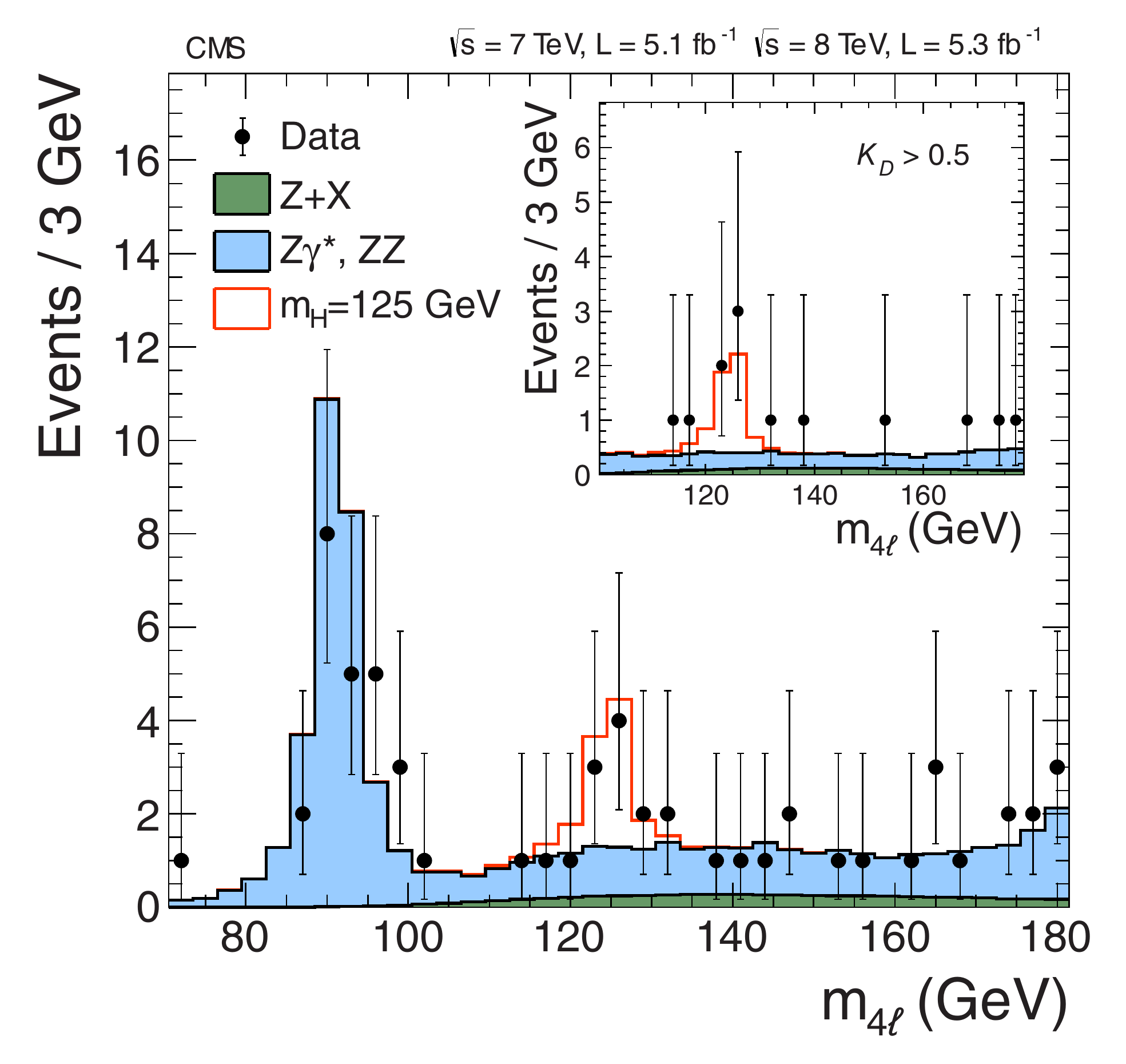}
	\end{minipage}}
	 \subfloat[]{%
	\begin{minipage}[c][1\width]{%
	   0.5\textwidth}
	   \centering%	
	   \includegraphics[width=1\textwidth]{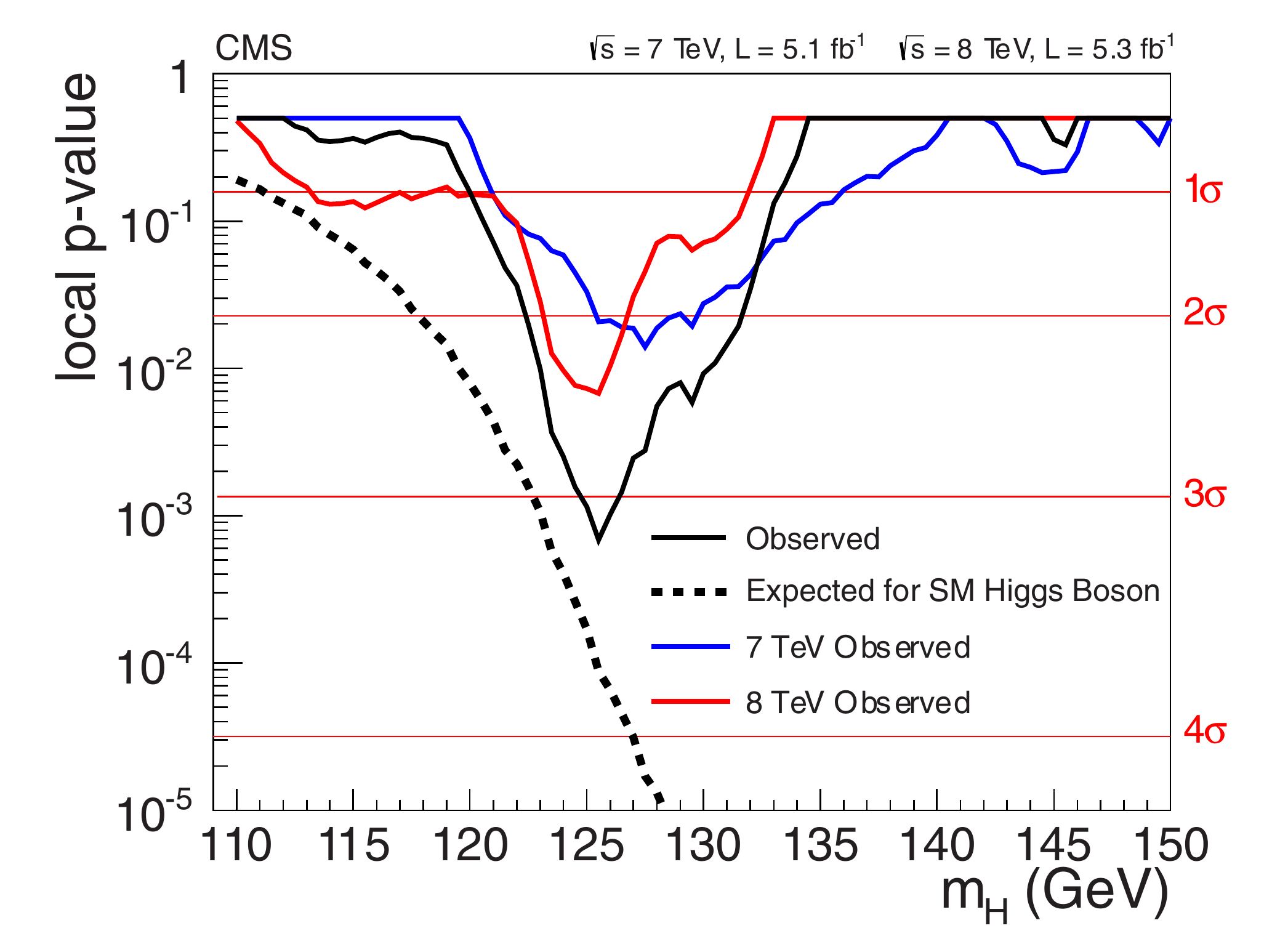}
	\end{minipage}}
     \caption{\small{ (a) Distribution of the four-lepton invariant mass
       for the $H \to ZZ\to 4\ell$  analysis. 
The points represent the data, the filled histograms represent the background, 
and the open histogram shows the signal expectation for a Higgs boson
of mass $m_H = 125$\,GeV, added to the background expectation.
The inset shows the $m_{4\ell}$ distribution after selection of events with
$K_D > 0.5$, as described in the text.
(b) The observed local $p$-value for the $ZZ$ decay mode as a function of the SM Higgs
boson mass. The dashed line shows the expected local $p$-values
for a SM Higgs boson with a mass $m_H$.
}}
\label{fig:ZZ}
\end{figure}

The $m_{4\ell}$ distribution is shown in Figure~\ref{fig:ZZ}(a).
There is a clear peak at the Z boson mass where the decay
$Z \to 4\ell$ is reconstructed.
This feature of the data is well reproduced by the background estimation.
The Figure~\ref{fig:ZZ}(a) also shows an excess of events above the expected
background around 125 GeV.
The kinematics of the $H \to ZZ\to 4\ell$ process in its centre-of-mass frame
is fully described by five angles and the invariant masses of 
the two lepton pairs~\cite{Cabibbo:1965zz,Gao:2010qx,DeRujula:2010ys}.
The variables provide significant discriminating power between signal and background.
A kinematic discriminant is constructed based on the probability ratio of the signal and background hypotheses,
$ K_{D} = \mathcal{P}_{\rm{sig}} / ( \mathcal{P}_{\rm{sig}} + \mathcal{P}_{\rm{bkg}} )$.
The likelihood ratio is defined for each value of $m_{4\ell}$.
The $m_{4\ell}$ distribution of events satisfying $K_D > 0.5$ is shown
in the inset in Figure~\ref{fig:ZZ}(a).
The statistical treatment requires, for each value of $m_{H}$, a
maximum likelihood fit of the two-dimensional distribution 
${\cal P}(m_{4\ell}, K_D | m_H) = {\cal P}(m_{4\ell} | m_H) 
\times {\cal P}(K_D | m_{4\ell})$.  
The fit is performed simultaneously in the 4e, 4$\mu$, and 2e2$\mu$ channels.

The expected 95\% CL upper limit on the signal strength
$\sigma/\sigma_\mathrm{SM}$,
in the background-only hypothesis,
for the combined 7~and 8 TeV
data, falls steeply between 110 and 140 GeV, and has a value
of 0.6 at $m_H = 125$ GeV.
The observed upper limit indicates the presence of a significant excess
in the range $120 < m_H < 130$ GeV. 
The local $p$-value is shown as a function of $m_H$ in Figure~\ref{fig:ZZ}(b) for the 7 and 8 TeV data, 
and for their combination.
The minimum local $p$-value in the data occurs at $m_H = 125.6$ GeV
and has a significance of 3.2$\,\sigma$ (expected 3.8$\,\sigma$).
The combined best-fit signal strength for a SM Higgs
boson mass hypothesis of 125.6 GeV is $\sigma/\sigma_\mathrm{SM} = 0.7^{+0.4}_{-0.3}$.

\subsection{$H \to WW\to \ell\nu\ell\nu$ decay mode} 

The decay mode $H \to WW\to \ell\nu\ell\nu$ is highly sensitive to a SM Higgs boson
in the mass range around the $WW$ threshold of 160 GeV~\cite{Chatrchyan:2012zz,Chatrchyan:2012ty}.
%%With the development of tools for lepton identification and $\ETmiss$
%%reconstruction optimized for LHC pileup conditions, it is
%%possible to extend the sensitivity down to 120\GeV.
This decay mode is analysed by selecting events in which both $W$ bosons decay leptonically, 
resulting in a signature with two isolated, oppositely charged leptons (electrons
or muons) and large missing transverse energy due to the undetected neutrinos.
%A $p_T$ threshold of 20 (10) GeV is applied to
%the lepton leading (subleading) in $p_T$.

Events are classified according to the number of jets (0, 1, or 2)
with $p_T >30$ GeV and within $|\eta|<4.7$ ($|\eta|<5.0$ for the 7 TeV
data set), and further separated into same-flavour ($ee$ and $\mu\mu$)
or different-flavour ($e\mu$) categories.
Events with more than two jets are rejected.  To improve the sensitivity of the analysis,
the selection criteria are optimized separately for the different
event categories since they are characterised by different dominating
backgrounds.  The zero-jet $e\mu$ category has the best signal
sensitivity.  Its main backgrounds are irreducible nonresonant $WW$
production and reducible $W$+jets processes, where a jet is
misidentified as a lepton.
The one-jet $e\mu$ and zero-jet same-flavour categories
only contribute to the signal sensitivity at the 10\% level because
of larger backgrounds, from top-quark decays and
Drell--Yan production, respectively.  Event selection in the two-jet
category is optimized for the VBF production mechanism.  This
category has the highest expected signal-to-background ratio, but its
contribution to the overall sensitivity is small owing to the
lower cross section relative to inclusive production.
Yields for the dominant backgrounds are estimated using control
regions in the data.  

\begin{figure}[htbp]
     \subfloat[]{%
	\begin{minipage}[c][1\width]{%
	   0.5\textwidth}
	   \centering%	
           \includegraphics[width=1\textwidth]{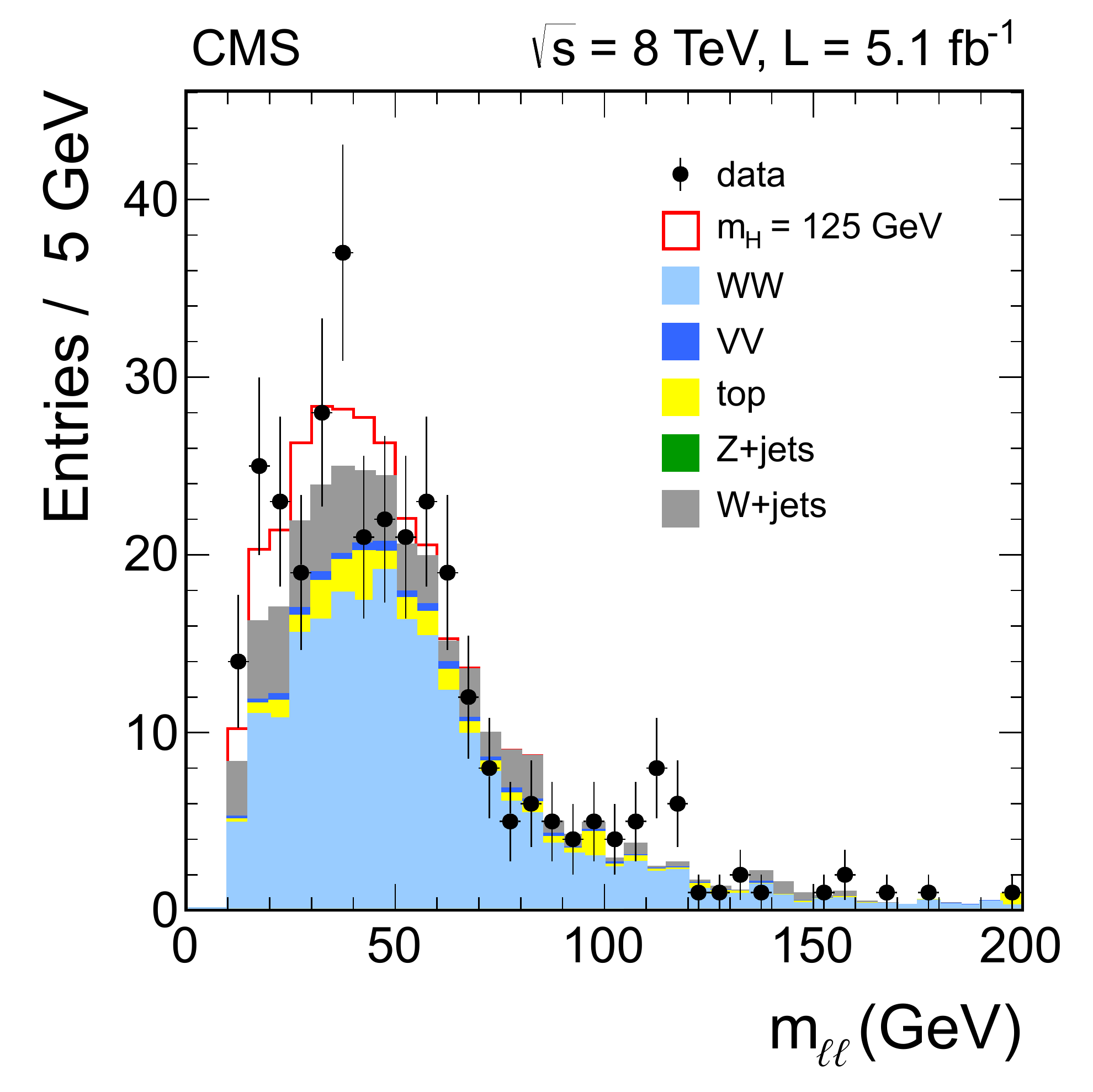}
	\end{minipage}}
	 \subfloat[]{%
	\begin{minipage}[c][1\width]{%
	   0.5\textwidth}
	   \centering%	
	   \includegraphics[width=1\textwidth]{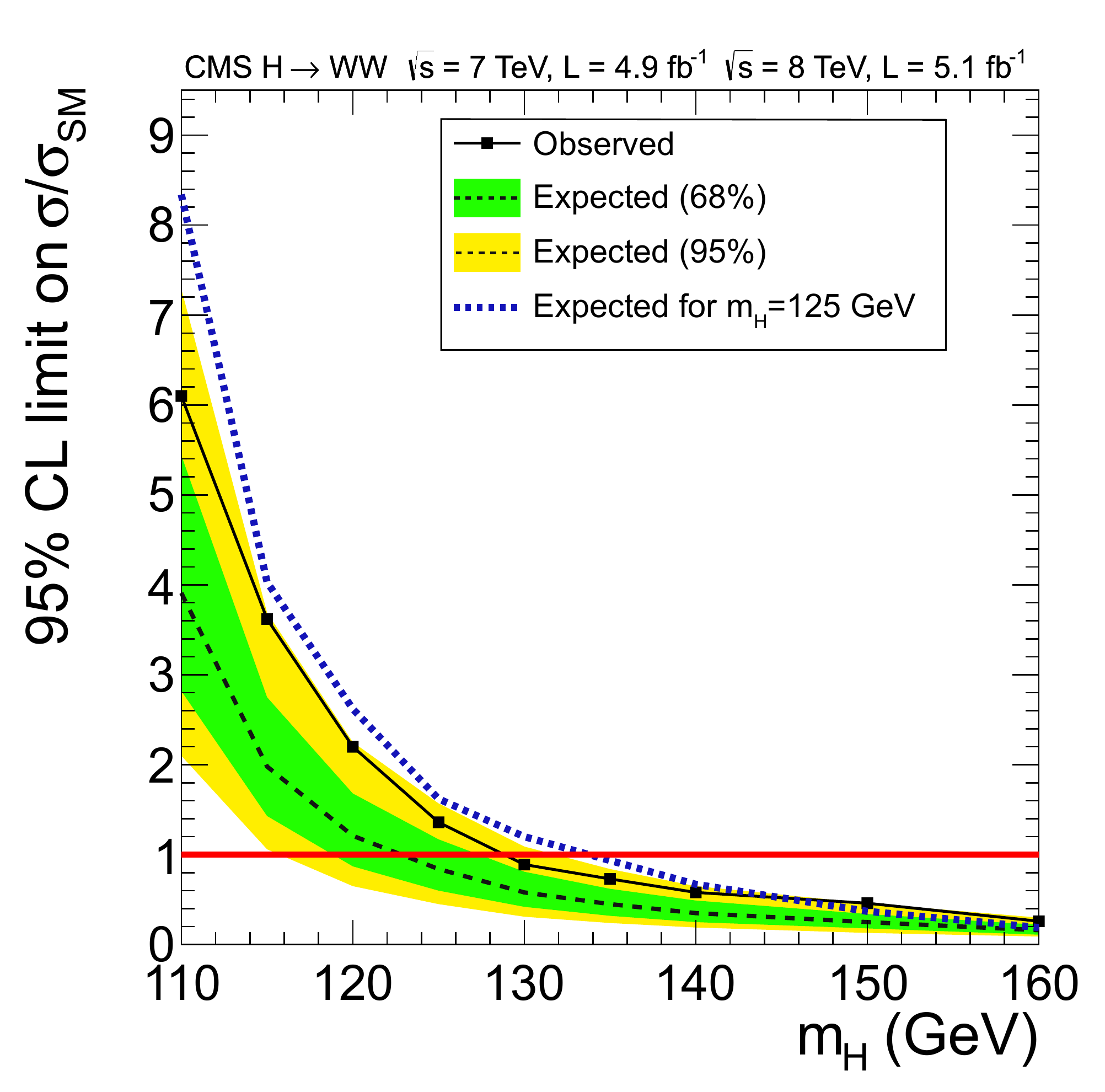}
	\end{minipage}}
     \caption{\small{ (a)Distribution of $m_{\ell\ell}$ for the
    zero-jet $e\mu$ category in the $H \to WW$ search
    at 8 TeV. The signal expected from a Higgs boson with a mass $m_H = 125$ GeV 
    is shown added to the background.
      decaying, via a W boson pair, to two leptons and two neutrinos,
      for the combined 7 and 8 TeV
      data sets. The symbol $\sigma/\sigma_\mathrm{SM}$ denotes the
      production cross section times the relevant branching fractions, relative to the SM expectation.
      The background-only expectations are represented by their median (dashed line) and by the 68\%  and 95\% CL bands. 
      The dotted curve shows the median expected limit for a SM Higgs boson with $m_H=125$ GeV.}}
    \label{fig:WW}
\end{figure}

One of the most sensitive variables to discriminate between
$H \to WW$ decays and nonresonant $WW$ production
is the dilepton invariant mass $m_{\ell\ell}$.
This quantity is shown in Figure~\ref{fig:WW}(a)
for the zero-jet $e\mu$ category after the full selection
for $m_H=125$ GeV, except for the
selection on $m_{\ell\ell}$ itself.
The 95\% CL expected and
observed limits for the combination of the 7 and 8 TeV
analyses are shown in Figure~\ref{fig:WW}(b).
A broad excess is observed that is consistent with a SM Higgs boson of mass 125 GeV.
This is illustrated by the dotted curve in Figure~\ref{fig:WW}(b) showing the median expected limit
in the presence of a SM Higgs boson with $m_H=125$ GeV.
The expected significance for a SM Higgs of mass
125 GeV is 2.4$\,\sigma$ and the observed significance is 1.6$\,\sigma$.

\subsection{$H \to bb$ decay mode} 

The $H \to bb$ analysis~\cite{Chatrchyan:2012zz,Chatrchyan:Hbb} selects events produced in association with a W or Z decaying via $Z \to ll, Z \to \nu\nu,$ or $W \to l\nu$. 
The selection requires two b-tagged jets and either two leptons, 
one lepton and missing transverse energy, or large missing transverse energy. 
The events are selected with a high transverse momentum of the di-jet system in order to improve the mass resolution 
and to suppress background processes. 
A multivariate classifier (BDT) is trained for different $m_H$ values and its output is used as the final discriminant.

Figure~\ref{fig:HttHbb}(b) shows as an example the BDT scores for the high-$p_T$ subchannel of the $Z(\nu\nu)H$ channel in the 8 TeV data set, 
after all selection criteria have been applied.
No significant deviation from the background expectation is observed.

\subsection{$H \to \tau\tau$ decay mode} 

The $H \to \tau\tau$ analysis~\cite{Chatrchyan:2012zz,Chatrchyan:2011nx} searches for a broad excess in the reconstructed di-tau invariant-mass distribution.
Events are classified according to the tau-lepton decay modes. 
Depending on the final state the event samples are further divided into exclusive subcategories, which are optimized for the sensitivity to a particular production mode. 
The category optimized for the VBF signature provides the largest sensitivity. 
Events failing these selection requirements are separated further, in a category with two jets optimized for associated production $HW(HZ)$, 
with one jet with large transverse momentum for associated production and production in gluon fusion, 
and with either no jets or with one with a small transverse momentum aiming at the gluon-fusion production process. 
The mass $m_{\tau\tau}$ is reconstructed with an algorithm~\cite{Chatrchyan:2011nx}  
combining the visible $\tau$ decay products and the missing transverse energy, achieving a resolution of about 20\% on $m_{\tau\tau}$.

\begin{figure}[htbp]
     \subfloat[]{%
	\begin{minipage}[c][1\width]{%
	   0.5\textwidth}
	   \centering%	
           \includegraphics[width=1\textwidth]{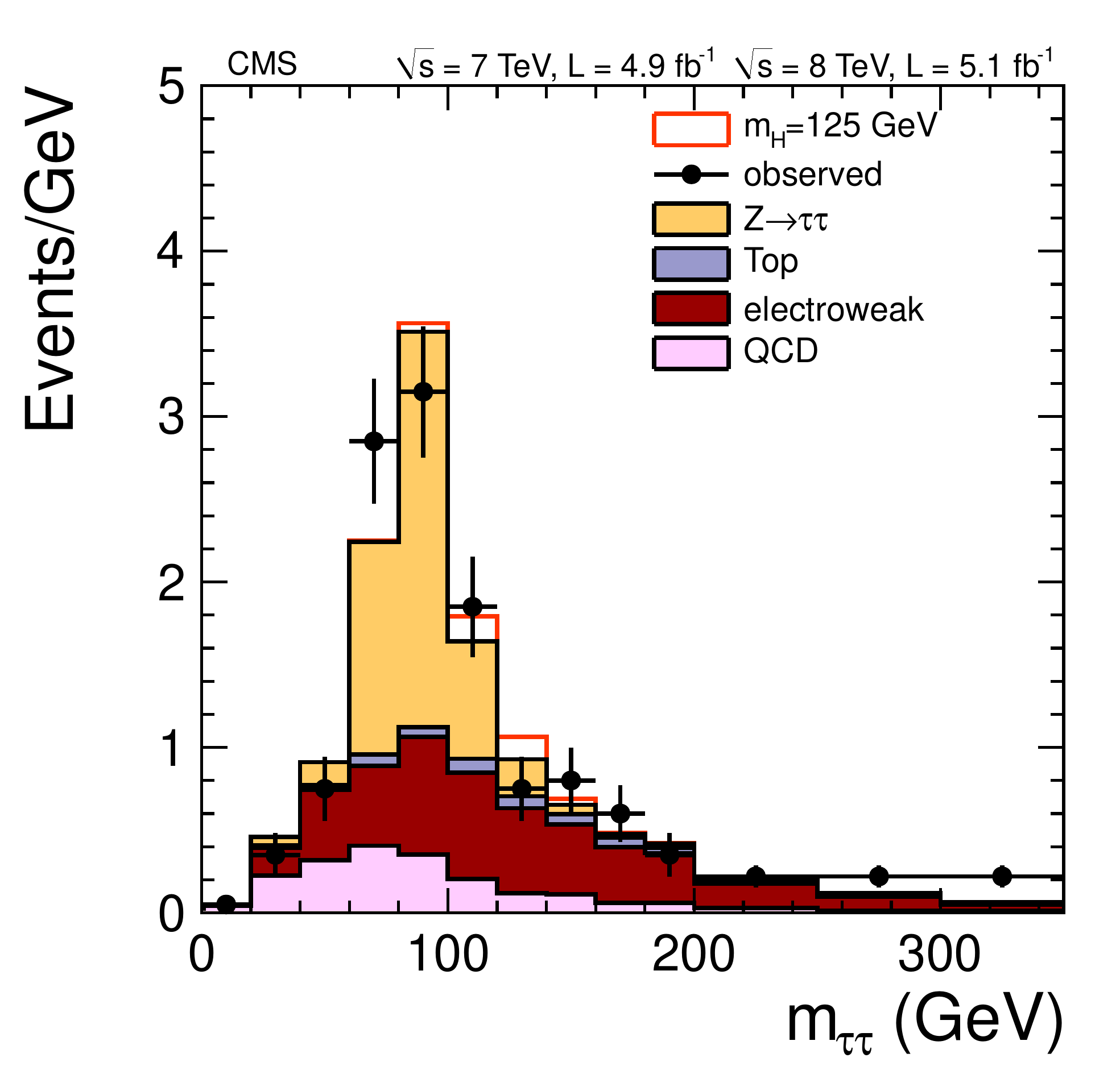}
	\end{minipage}}
	 \subfloat[]{%
	\begin{minipage}[c][1\width]{%
	   0.5\textwidth}
	   \centering%	
	   \includegraphics[width=1\textwidth]{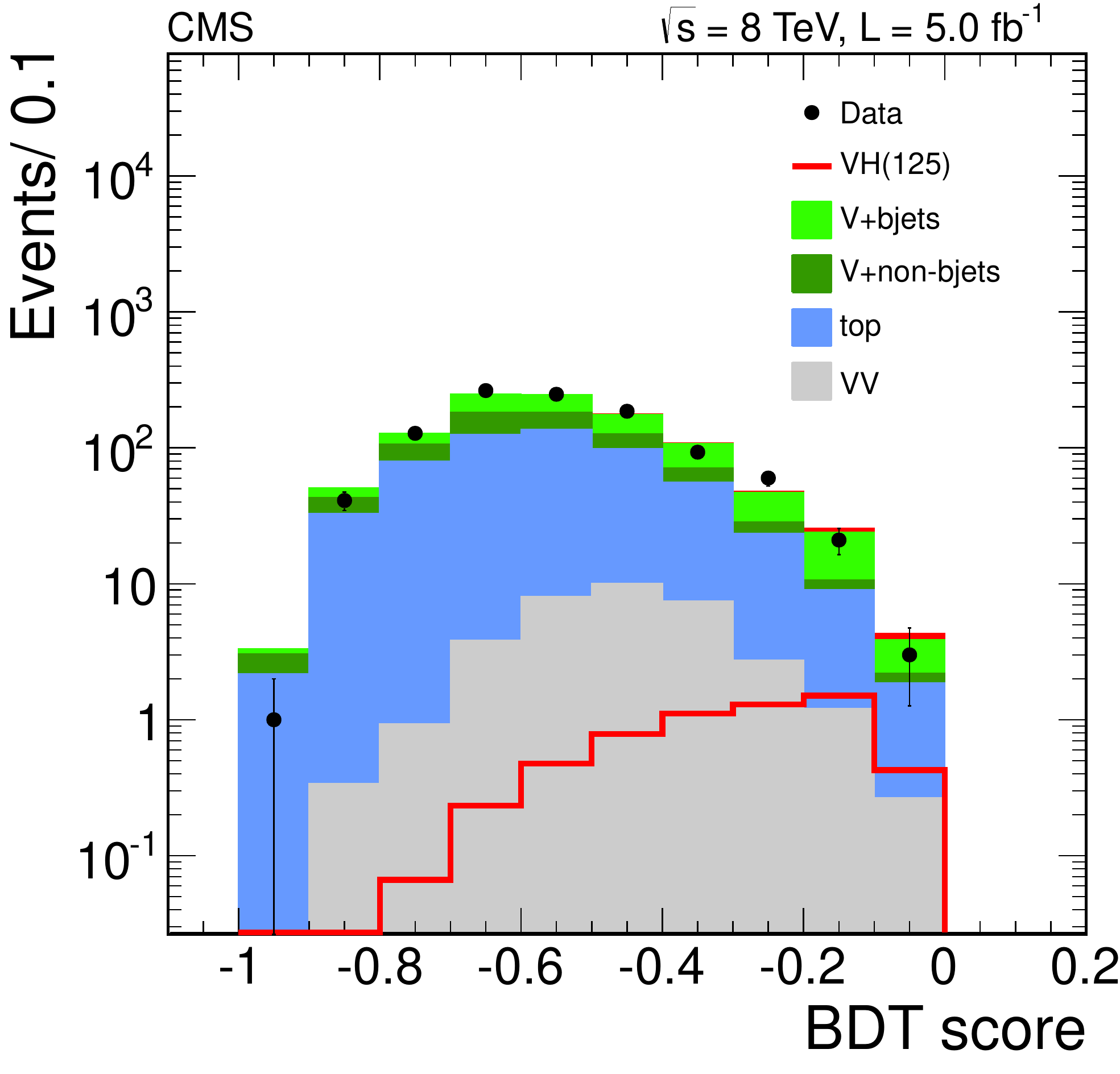}
	\end{minipage}}
     \caption{\small{ (a) Distribution of $m_{\tau\tau}$ in the combined 7~and 8 TeV data sets for the $\mu\tau_{\mathrm h}$ VBF category of the
    $ H \to \tau\tau$ search. The signal expected from a SM Higgs boson
    ($m_H = 125$ GeV) is added to the background.
    (b) Distribution of BDT scores for the high-$p_T$ subchannel of the  $Z(\nu\nu)H(bb)$ search in the 8 TeV data set
     after all selection  criteria have been applied.
     The signal expected from a Higgs boson ($m_H = 125$ GeV)
     is shown added to the background and also overlaid for comparison with the diboson background.
     }}
    \label{fig:HttHbb}
\end{figure}

The main irreducible background arises from $Z \to \tau\tau$ production, 
whose di-tau invariant-mass distribution is derived from data by selecting $Z \to \mu\mu$ events, 
in which the reconstructed muons are replaced with reconstructed particles from the decay of simulated $\tau$ leptons of the same momenta. 
The reducible backgrounds from $W$+jets, Drell--Yan, and multi-jet production are also evaluated from control samples in data. 
The distribution of the di-tau mass distribution is shown in Figure~\ref{fig:HttHbb}(a). 
No significant deviation from the background expectation is observed.

\section{Observation and anatomy of a new particle}

The individual results for the channels analysed for the five decay
modes, summarised in Table~\ref{tab:chans}, are combined using the
methods outlined here~\cite{LHC-HCG-Report}.
The combination assumes the relative branching fractions predicted by the SM 
and takes into account the experimental statistical and
systematic uncertainties as well as the theoretical uncertainties, which are
dominated by the imperfect knowledge of the QCD scale and parton distribution functions.
The $CLs$ method is used to compute the exclusion limit.
The median expected exclusion range of $m_H$ at 95\% CL in the absence of a signal is 110-600 GeV. 
In most of the explored Higgs boson mass range, the differences between the observed and expected limits 
are consistent with statistical fluctuations since the observed limits are generally within the $1\sigma$ and $2\sigma$ 
bands of the expected limit values. 
However at low mass, in the range 122.5 $< m_H <$ 127 GeV, 
an excess of events is observed which makes the observed limits considerably weaker than expected in the absence of a SM Higgs boson 
and, hence, does not allow exclusion.

\begin{figure}[htbp]
     \subfloat[]{%
	\begin{minipage}[c][1\width]{%
	   0.5\textwidth}
	   \centering%	
           \includegraphics[width=1\textwidth]{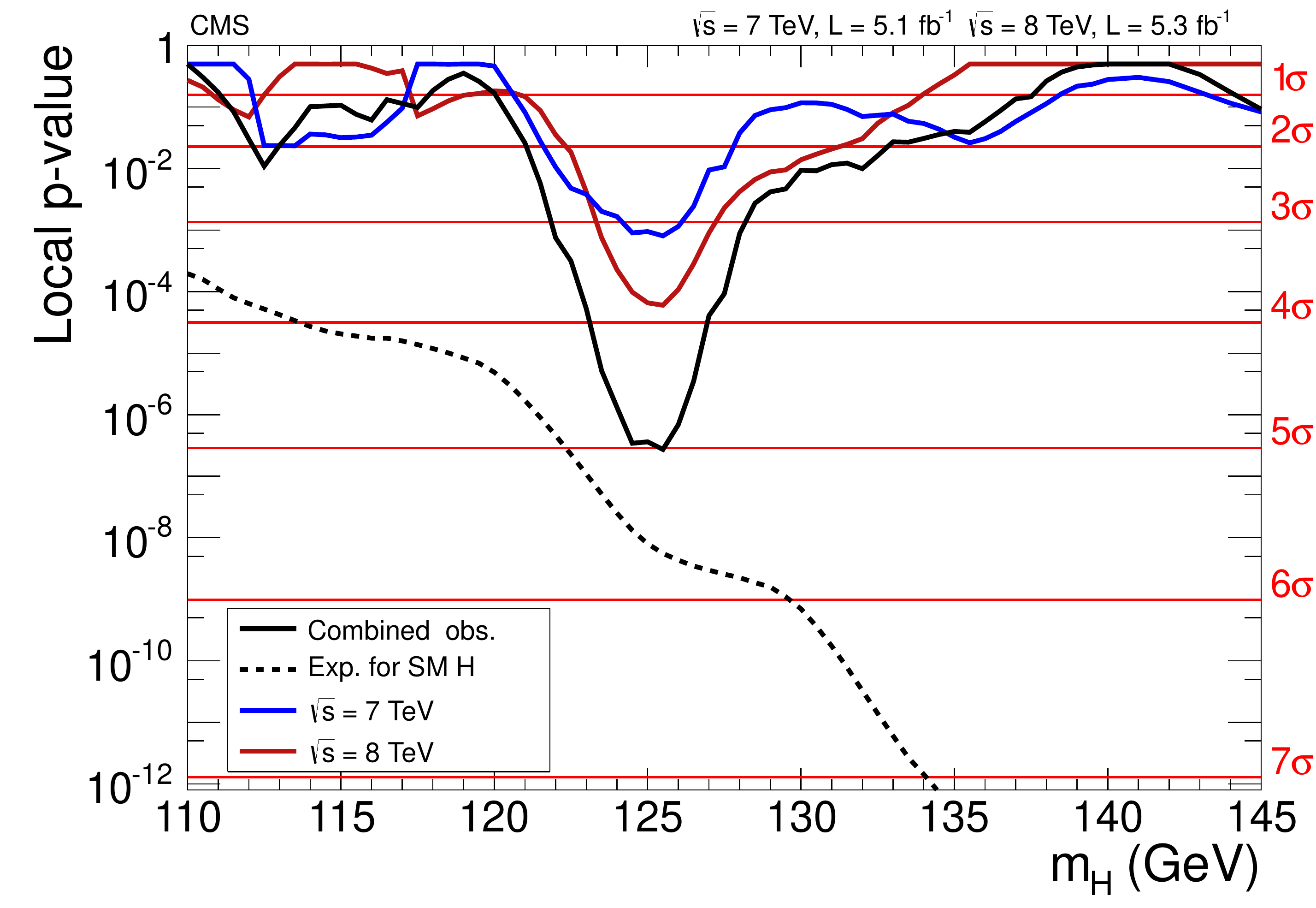}
	\end{minipage}}
	 \subfloat[]{%
	\begin{minipage}[c][1\width]{%
	   0.5\textwidth}
	   \centering%	
	   \includegraphics[width=1\textwidth]{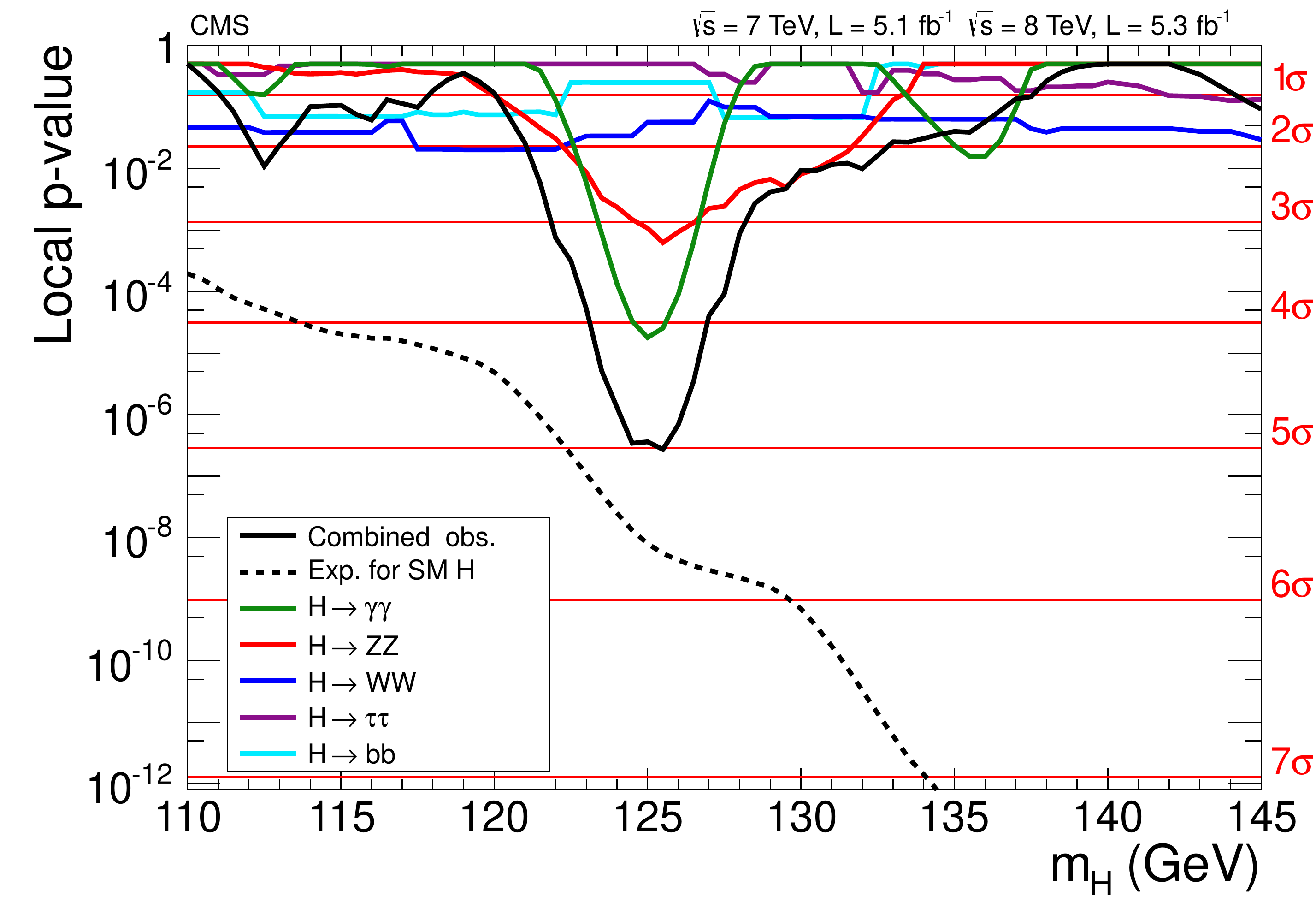}
	\end{minipage}}
     \caption{ \small{
    (a) The observed local $p$-value for 7 TeV~and 8 TeV data, and their combination
    as a function of the SM Higgs boson mass. The dashed line shows the expected local $p$-values
    for a SM Higgs boson with a mass $m_H$.
    (b) The observed local $p$-value for the five decay modes and the overall combination as a function of the SM Higgs boson mass.
    The dashed line shows the expected local $p$-values
    for a SM Higgs boson with a mass $m_H$. 
    }}
    \label{fig:Obs}
\end{figure}

The consistency of the observed excess with the background-only hypothesis may be judged from
Figure~\ref{fig:Obs}(a), which shows a scan of the
local $p$-value for the 7~and 8 TeV data sets
and their combination.
In the overall combination the significance is 5.0\,$\sigma$ for $m_H=125.5$ GeV.
Figure~\ref{fig:Obs}(b) gives the local $p$-value for the five
decay modes individually and displays the expected overall $p$-value.
The largest contributors to the overall excess in the combination
are the $\gamma\gamma$ and $ZZ$ decay modes. 
They both have very good
mass resolution, allowing good localization of the invariant mass of
a putative resonance responsible for the excess. 
Table~\ref{tab:Signif} summarises the expected and observed local $p$-values for a SM Higgs boson mass
hypothesis of 125.5 GeV for the various combinations of channels.

\begin{table}[htbp]
\begin{center}
\caption{
The expected and observed local $p$-values, expressed
as the corresponding number of standard deviations of the observed excess from the
background-only hypothesis,
for $m_H = 125.5$ GeV, for various combinations of decay modes.
}
\label{tab:Signif}
\begin{tabular}{l|c|c}
\hline
Decay mode/combination & Expected ($\sigma$) & Observed ($\sigma$) \\
\hline\hline
$\gamma\gamma$ & 2.8 & 4.1 \\ %
$ZZ$  &  3.8 & 3.2 \\ %
\hline
$\tau\tau$ + $bb$ & 2.4 & 0.5 \\
$\gamma\gamma$ + $ZZ$ & 4.7 & 5.0 \\
$\gamma\gamma$ + $ZZ$ + $WW$ & 5.2 & 5.1 \\ %
% \hline
$\gamma\gamma$ + $ZZ$ + $WW$ + $\tau\tau$ + $bb$ & 5.8 & 5.0 \\%
\hline
\end{tabular}
\end{center}
\end{table}

\begin{figure} [htbp]
\begin{center}
\includegraphics[width=0.5\textwidth]{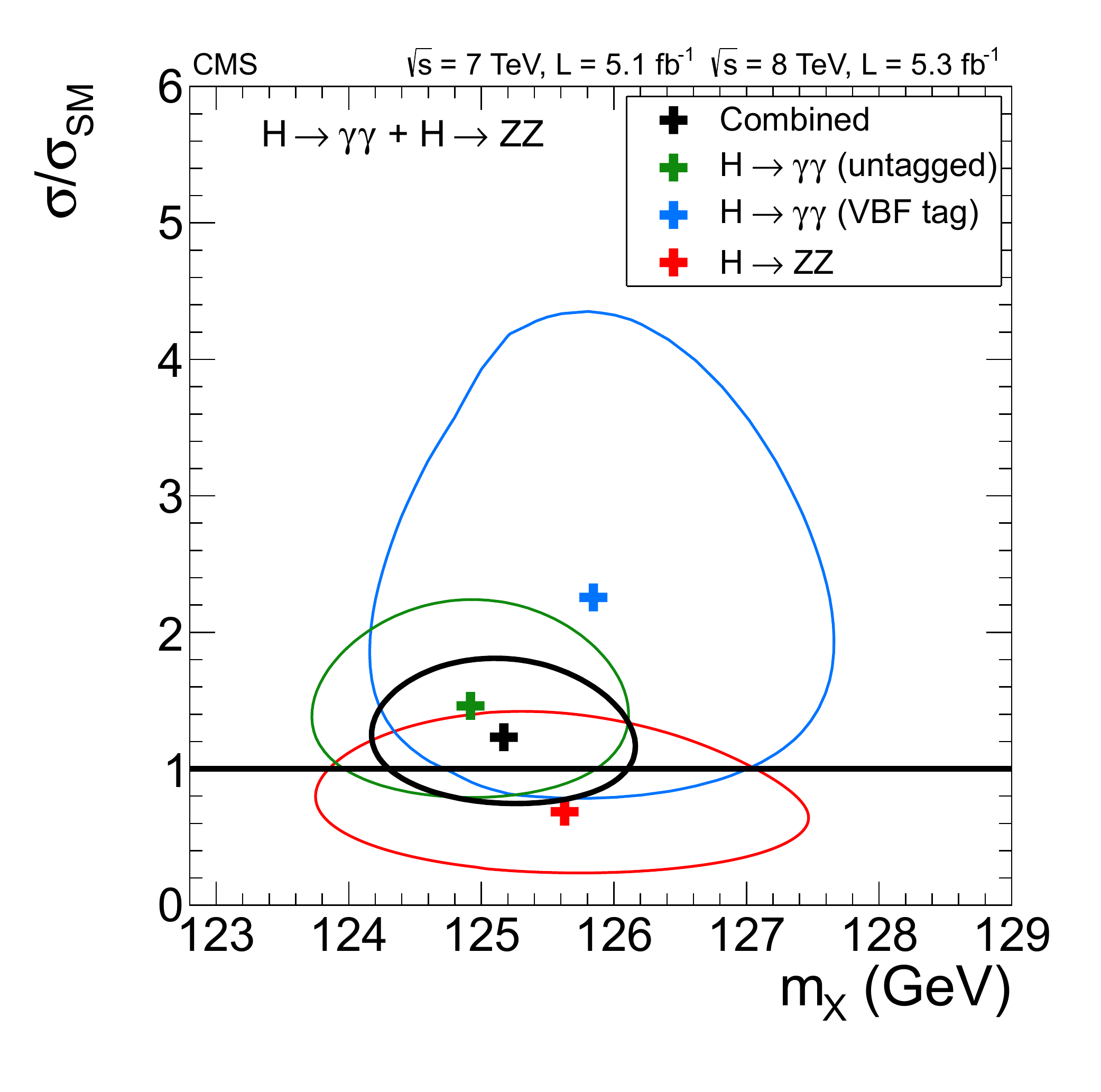}
\caption{\small{
The 68\% CL contours for the signal strength $\sigma/\sigma_{\mathrm{SM}}$ 
versus the boson mass $m_{X}$ for the untagged $\gamma \gamma$,
$\gamma \gamma$ with VBF-like dijet, 4$\ell$, and their combination. 
The symbol $\sigma/\sigma_{\mathrm{SM}}$ denotes the 
production cross section times the relevant branching fractions,
relative to the SM expectation.In this combination, the relative signal strengths for the three
decay modes are constrained by the expectations for the SM Higgs boson.}}
\label{fig:fit_mass}
\end{center}
\end{figure}

%mass
To measure the mass of the observed state, we use the $\gamma\gamma$ and $ZZ \to 4\ell$ channels 
that have excellent mass resolution and for which we observe the excess with a high significance.
Figure~\ref{fig:fit_mass} shows 2D 68\% 
confidence level regions for two parameters of interest, the signal strength $\mu$ and mass $m_X$ for the three channels 
(untagged $\gamma\gamma$, VBF-tagged $\gamma\gamma$, and $ZZ \to 4l$). 
The three channels are consistent and thus can be combined.
The combined best-fit mass is $m_X = 125.3 \pm 0.4 \rm{(stat)} \pm 0.5 \rm{(syst)}$ GeV.

\begin{figure}[htbp]
     \subfloat[]{%
	\begin{minipage}[c][1\width]{%
	   0.5\textwidth}
	   \centering%	
           \includegraphics[width=1\textwidth]{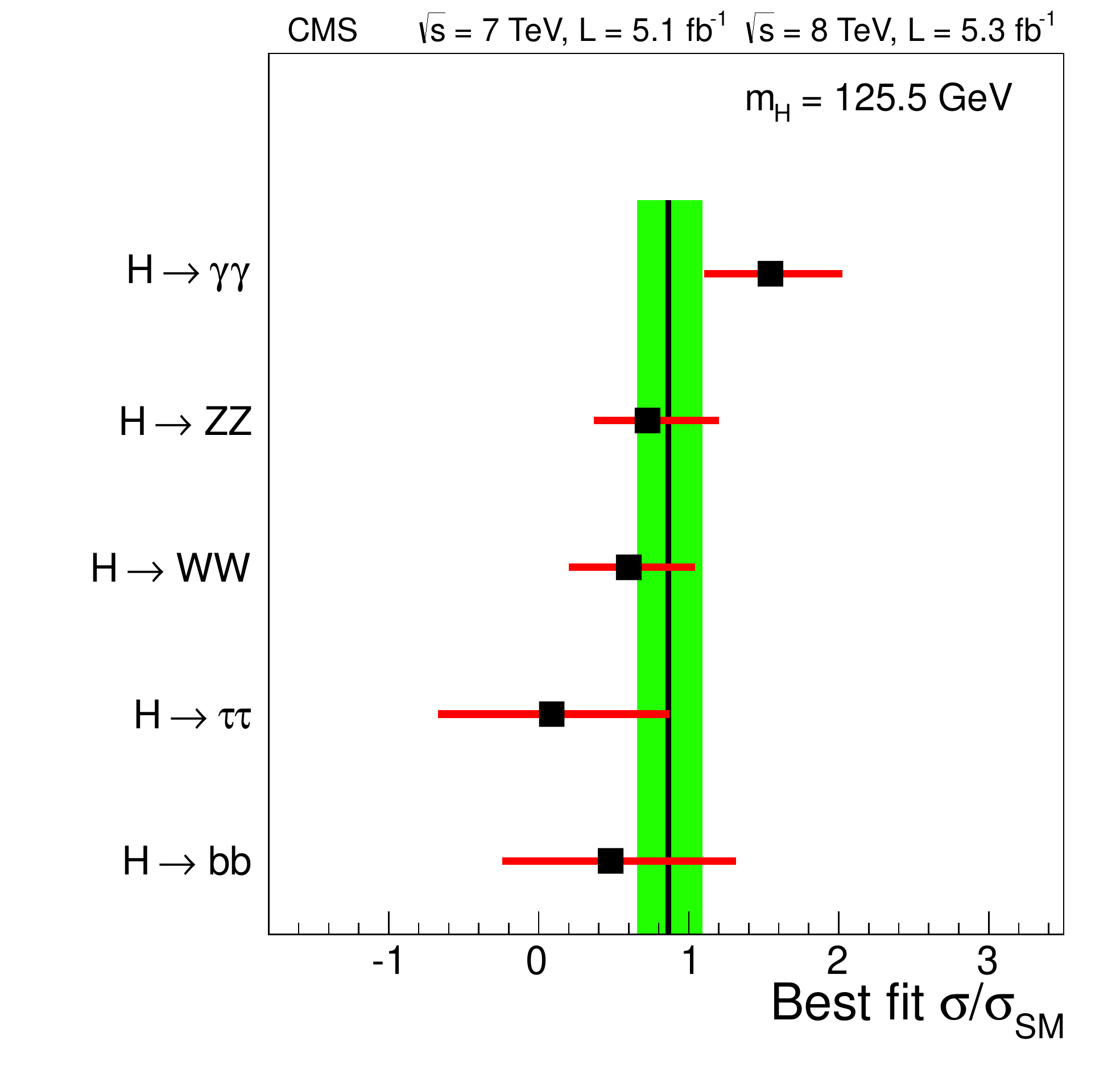}
	\end{minipage}}
	 \subfloat[]{%
	\begin{minipage}[c][1\width]{%
	   0.5\textwidth}
	   \centering%	
	   \includegraphics[width=1\textwidth]{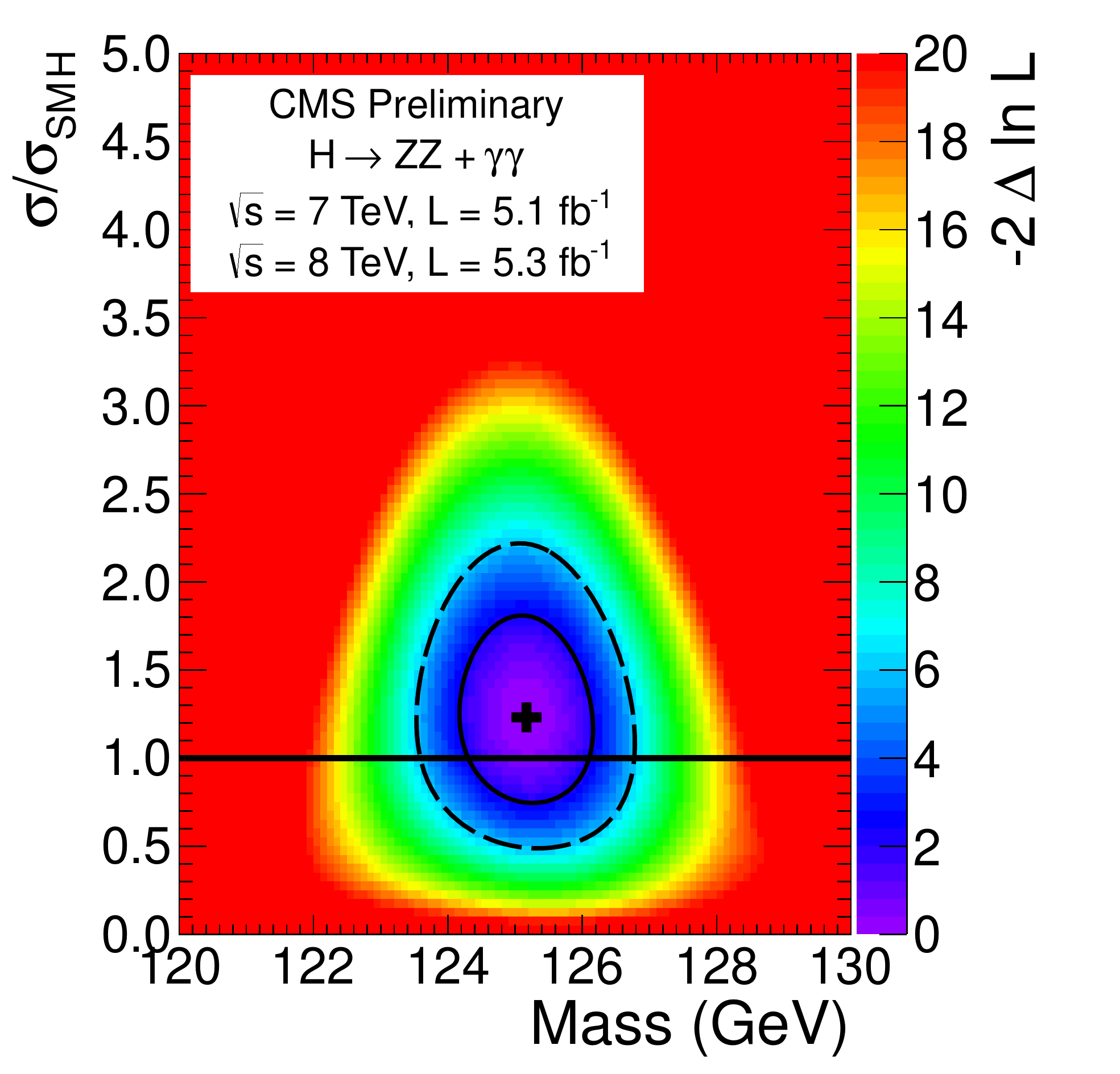}
	\end{minipage}}
     \caption{ \small{
    (a) Values of $\sigma/\sigma_{\mathrm{SM}}$ for the combination (solid vertical line)
    and for individual decay modes (points).
    %The vertical band shows the overall $\sigma/\sigma_\text{SM}$ value \MUHAT.
    The symbol $\sigma/\sigma_{\mathrm{SM}}$ denotes the
    production cross section times the relevant branching fractions, relative to the SM expectation.
    The horizontal bars indicate the $\pm 1$ standard deviation uncertainties
    in the $\sigma/\sigma_{\mathrm{SM}}$ values for individual modes, statistical and systematic uncertainties
    are included.
    (b) The	2D-scan test statistic	 -2ln(Q) vs the ($c_V$, $c_F$) parameters.	
    The cross indicates the best-fit values. 
    The solid and dashed contours show the 68\% and 95\% CL ranges, respectively. 
    The yellow diamond shows the SM point ($c_V$, $c_F$)=(1,1).
     }}
    \label{fig:comp}
\end{figure}

%The p-value characterises the probability of background producing an observed excess of events, 
%but it does not give information about the compatibility of an excess with an expected signal.
We then performed a limited number of
compatibility test with the SM Higgs boson hypothesis.
A first test of the compatibility of the observed boson with the SM
Higgs boson is provided by examination of
the best-fit value for the signal strength relative to the SM expectation, $\sigma/\sigma_{\mathrm{SM}}$.
Figure~\ref{fig:comp}(a) shows
the values of $\sigma/\sigma_{\mathrm{SM}}$ for individual decay modes and their combination.  
The horizontal bars indicate the $\pm 1$ standard deviation uncertainties,
statistical and systematic uncertainties are included.
Electroweak symmetry breaking via the Higgs mechanism sets a well-defined ratio for the
couplings of the Higgs boson to the W and Z bosons, $g_H^{WW}/g_H^{ZZ}$, 
protected by the custodial symmetry.
To quantify such consistency, we introduce two event rate modifiers $\mu_{ZZ}$ and $R_{wz}$. 
The expected $H \to ZZ \to 4\ell$ event yield is scaled by $\mu_{ZZ}$, 
while the expected untagged  $H \to WW \to \ell\nu\ell\nu$ event yield is scaled
by $R_{wz} \times \mu_{ZZ}$. 
A scan of the test statistic q($R_{wz}$), 
while profiling all other nuisances and the signal strength
modifier $\mu_{ZZ}$, yields $R_{wz} = 0.9 ^{+1.1}_{-0.6}$.
Given that $R_{wz}$ is consistent with unity, we now assume that the ratio of the couplings of the observed state 
to the W and Z bosons is as required by the custodial symmetry. 
Under this assumption, we can check the compatibility of the observation with the standard model Higgs boson 
by introducing and fitting for two free parameters $c_V$ and $c_F$. 
The first, $c_V$, scales the standard model Higgs boson couplings to the W and Z bosons, 
while preserving their ratio. 
The other, $c_F$, scales all couplings to fermions by one constant factor. 
The 2D likelihood scan and the 68\% and 95\% confidence regions for $c_V$ and $c_F$ are shown in Figure\ref{fig:comp}(b).
In this scan $c_V$ and $c_F$ are constrained to be positive. 
The data are compatible with the expectation for the SM Higgs boson,
the point ($c_V$, $c_F$)=(1,1) is within the 95\% confidence interval defined by data. 
%These results are consistent, within uncertainties, with the
%expectations for a SM Higgs boson.

\section{Searches beyond standard model Higgs boson}

\begin{figure}[htbp]
     \subfloat[]{%
	\begin{minipage}[c][1\width]{%
	   0.5\textwidth}
	   \centering%	
           \includegraphics[width=1\textwidth]{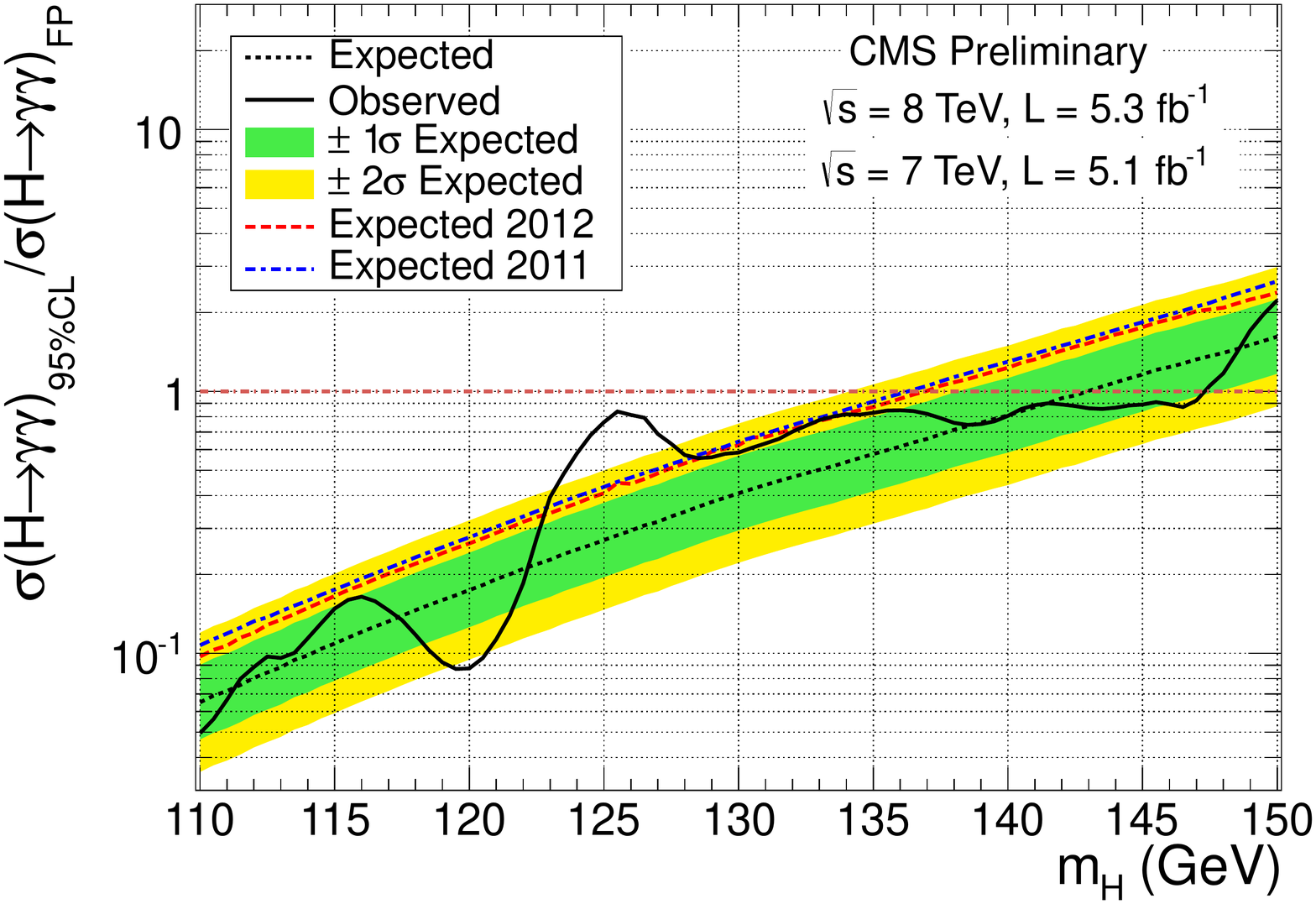}
	\end{minipage}}
	 \subfloat[]{%
	\begin{minipage}[c][1\width]{%
	   0.5\textwidth}
	   \centering%	
	   \includegraphics[width=1\textwidth]{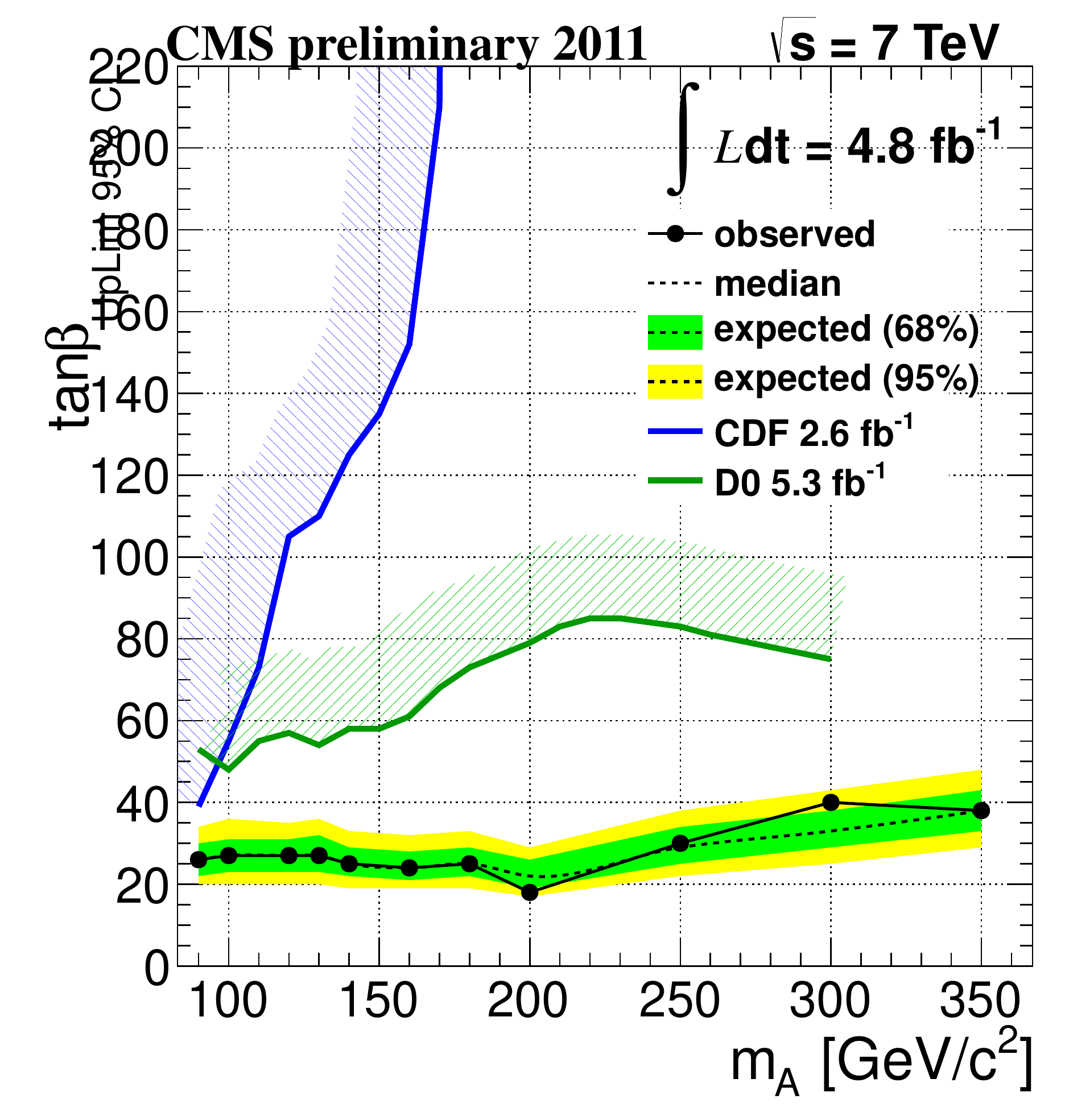}
	\end{minipage}}
     \caption{ \small{
    (a) The combined 2011 and 2012 exclusion limit on the cross section of a FP Higgs boson 
decaying into two photons as a function of the boson mass relative to the FP cross section.    
   (b) For the search of neutral MSSM Higgs boson produced in association with two spectator $b$-quarks 
and decaying semi-leptonical to pairs of $b$-quarks
observed and expected upper limit with 95\% confidence level in the MSSM plane ($M_A$ , tan$\beta$), 
for $m_h^{max}$ benchmark scenario, with $\mu = -200$ GeV, 
overlaid to the previously published CDF and D0 results.
     }}
    \label{fig:MSSM}
\end{figure}

The CMS collaboration is strongly active in searches for beyond SM Higgs boson.
At the moment there is no evidence for any excess above backgrounds
but strong constraints are imposed.
In following a brief summary is presented.

In an simple extension of the SM two models have been tested.
The first one is the SM including a fourth generation of fermions 
(the SM4 model)~\cite{CMS-PAS-HIG-12-008} 
where additional heavy quarks in the quark loop associated with the $gg \to H$ 
process greatly enhance its production cross-section while
other production mechanisms are not affected.
The second one includes a fermiophobic (FP) Higgs boson~\cite{CMS-PAS-HIG-12-008,CMS-PAS-HIG-12-022} where 
a Higgs boson couples only to the vector bosons at tree level. 
In such a model the branching fraction for a low mass FP Higgs boson 
to decay to two photons is enhanced by an order of magnitude with respect to the SM.
For example Figure~\ref{fig:MSSM}(a) 
shows the combined 2011 and 2012 exclusion limits on the cross section of a FP Higgs boson 
decaying into two photons as a function of the boson mass relative to the FP cross section.

Analyses have been performed  in the contest both of 
the Minimal Supersymmetric Standard Model (MSSM) with two Higgs doublets~\cite{Chatrchyan:2011nx,CMS-PAS-HIG-12-026,CMS-PAS-HIG-12-027,CMS-PAS-HIG-12-033}
and of the Next-to-Minimal Supersymmetric Standard Model (nMSSM) with additional scalar field~\cite{CMS-PAS-HIG-12-004}.
In the search of neutral MSSM Higgs boson produced in association with two spectator $b$-quarks 
and decaying semi-leptonical to pairs of $b$-quarks,
Figure~\ref{fig:MSSM}(b) shows the
observed and expected upper limits with 95\% confidence level in the MSSM plane ($M_A$ , tan$\beta$), 
for $m_h^{max}$ benchmark scenario, with $\mu = -200$ GeV, 
overlaid to the previously published CDF and D0 results.

Searches have been performed in the 
extension of the SM which include the seesaw mechanism of type II, 
this allows the existent of a doubly charged Higgs boson~\cite{CMS-PAS-HIG-12-005}.
The doubly charged Higgs boson 
is excluded in mass ranges significantly beyond those explored previously by the LEP and Tevatron experiments.

\section{Conclusion} 

Results are presented from searches for the SM Higgs boson in $pp$ collisions
at $\sqrt{s} = 7$ and 8 TeV in the CMS experiment at the LHC.
An excess of events is observed above the expected background, with a local significance of 5.0$\sigma$, 
at a mass near 125 GeV, signalling the production of a new particle.
The results are consistent, within uncertainties, with the
expectations for a SM Higgs boson.
The searches beyond SM Higgs boson are highlighted and 
at the moment there is no evidence for any excess above backgrounds but strong constraints are imposed.

\end{document}